\PassOptionsToPackage{colorlinks=true,linkcolor=blue,citecolor=blue,urlcolor=blue}{hyperref}
\documentclass[prb,twocolumn,superscriptaddress,floatfix,longbibliography]{revtex4-2}

\usepackage{graphicx}
\usepackage{amsmath}
\usepackage{amssymb}
\usepackage{xcolor}
\usepackage{hyperref}

	\hypersetup{
		pdftitle={Chirally pinned flat bands, helical minibands, and bias-controlled topology of Jackiw–Rebbi states},
		pdfsubject={Topological bound states in a two-orbital model with a periodically modulated inter-orbital hybridization},
		pdfkeywords={Jackiw-Rebbi, two-orbital model, inter-orbital hybridization, flat band, helical miniband, SSH model, cold atoms}
	}
	
\begin{document}
		
		\title{Chirally pinned flat bands, helical minibands, and bias-controlled topology of Jackiw–Rebbi states in a periodically modulated Dirac mass texture}
		
		\author{Tao Zhou}
		\email{tzhou@scnu.edu.cn}
		\affiliation{Guangdong Basic Research Center of Excellence for Structure and Fundamental Interactions of Matter, Guangdong Provincial Key Laboratory of Quantum Engineering and Quantum Materials, School of Physics, South China Normal University, Guangzhou 510006, China}
		\affiliation{Guangdong-Hong Kong Joint Laboratory of Quantum Matter, Frontier Research Institute for Physics, South China Normal University, Guangzhou 510006, China}

		\begin{abstract}
			Flat bands amplify interactions into correlated phases, and
			topological bound states are robust by principle; combining the two in
			a model that is both exactly analyzable and experimentally accessible
			is therefore especially desirable. Here we show that a square-lattice
			model with a slowly modulated Dirac mass texture,
			$m(x)=m+m_0\cos(2\pi x/L)$ with $L$ up to hundreds of sites, provides
			such a setting, without engineered geometries, lattice-scale textures,
			or long-range hoppings. Exact nesting reduces the problem,
			independently at each transverse momentum $k_y$, to a $1+1$
			dimensional Dirac Hamiltonian with a position-dependent mass: every
			simple zero of $m(x)$ binds Jackiw--Rebbi zero modes, and the two
			walls of a period form an effective Su--Schrieffer--Heeger ring. The
			model interpolates between a chirally pinned flat band at large $L$, a
			helical miniband whose nodes are enforced by a boundary phase of the
			ring, and a bias-dimerized regime in which the sign of the bias sets a
			winding number. A central quantitative result is that the apparent
			bandwidth of this miniband is not a single scale: a max--min
			measurement conflates three widths of different origin: the helical
			height $W_{\rm JR}$, exponential in the period; the zone-edge upturn
			$W_{\rm edge}\propto L^{-2/3}$, which dominates at large $L$; and a
			bias bump, exponential in $L$, so the flat band is robust against bias
			in the thermodynamic limit. We analyze two lattice realizations, a
			two-orbital model with a modulated inter-orbital hybridization and a
			single-orbital bond-order texture that shares the nodal physics
			while leaving the van Hove points gapless, and propose experiments
			on cold-atom superlattices, Kekul\'e-ordered graphene, STM designer
			lattices, and microwave arrays. The construction is analytically
			tractable and each control parameter has a direct experimental
			counterpart, providing a reference model for flat-band and
			domain-wall physics.
		\end{abstract}
		
		\maketitle
		
		\section{Introduction}
		\label{sec:intro}
		
		Domain-wall bound states are among the oldest and most robust
		consequences of relativistic band structures in condensed matter. The
		Jackiw--Rebbi mechanism \cite{jackiw1976} binds a zero-energy state to
		any sign change of a Dirac mass, and its solid-state realization as
		solitons in polyacetylene \cite{su1979,jackiw1981,rice1982} established
		the modern language of topological band theory. When the mass pattern is
		made periodic, the individual zero modes hybridize into bands, and the
		question of what controls their dispersion becomes central: flat
		zero-mode bands promise strongly correlated physics
		\cite{cao2018,andrei2021,leykam2018}, while dispersive ones offer
		controllable transport channels.

		Mass superlattices have been studied extensively in graphene-like systems, where a spatially modulated mass or potential creates Jackiw--Rebbi-like modes that disperse along the walls \cite{semenoff2008,demartino2023,anand2025}. In those systems the bound-state dispersion is set by the momentum parallel to the wall and is generically chiral, so the modes carry current.

		The existing mass-superlattice and flat-band literature leaves two issues unresolved. First, while the wall-parallel dispersion of Jackiw--Rebbi modes is well established~\cite{semenoff2008,demartino2023,anand2025}, it is not known whether transverse momentum can instead enter zero-mode interference as a boundary phase of an effective Su--Schrieffer--Heeger (SSH) ring, so that the zero modes form a flat or helical miniband whose dispersion is governed by transverse interference rather than by chiral propagation along the walls. 
		Second, most two-dimensional flat-band constructions rely on special lattice geometries or fine-tuned long-range hoppings~\cite{leykam2018}. Whether a minimal square-lattice model with only nearest-neighbor hopping and a modulated on-site hybridization can produce a symmetry-pinned flat band, and whether a uniform bias can drive it into a dimerized topological regime, remain open. 
		
		These questions are worth answering in a concrete, tunable setting.
		A model
		that combines flat bands with domain-wall bound states, and in which
		flat-band formation, domain-wall hybridization, and topological
		switching are controlled by independent parameters, would provide a common
		platform for strongly correlated physics and topological transport.
		A second motivation is diagnostic. The bandwidth of a zero-mode
		band is the quantity most directly read off from data, and if it mixes
		contributions with different period dependences, hybridization scales
		inferred from it are unreliable; a model in which each contribution is
		separately computable provides the reference point for interpreting
		such measurements. The construction analyzed below is promising in
		both respects: it uses only standard square-lattice ingredients, and
		each of its control parameters (the period, the modulation amplitude,
		and the bias) has a direct experimental counterpart in cold-atom,
		graphene, STM-designer, and photonic platforms.
		
		In this paper, we consider a two-orbital model on the square lattice: two orbitals with opposite-sign nearest-neighbor hopping, $\epsilon_b(\mathbf k)=-\epsilon_a(\mathbf k)$, coupled by an on-site inter-orbital hybridization whose amplitude is slowly modulated along one direction. The modulation is periodic, with a tunable uniform offset, so that the mass texture has simple, well-separated zeros and can be biased continuously. The 
		opposite-sign dispersions are exactly nested --- the two orbitals play the role of the two square-lattice valleys --- so the problem reduces, independently at each transverse momentum 
		$k_y$, to an effective one-dimensional Dirac Hamiltonian with a position-dependent mass. Every simple zero of the mass texture binds Jackiw--Rebbi zero modes, and the two walls of a period form an effective SSH ring. On the lattice, the zero-energy pinning is protected by an orbital-interchange chiral symmetry, which is exact for any on-site hybridization texture.

		The model exhibits three regimes: a chirally pinned flat zero-energy band at large periods, a helical miniband whose nodes are enforced by a boundary phase of the effective SSH ring, and a bias-dimerized regime in which the sign of the bias sets a winding number. A central quantitative result is that the apparent bandwidth of the central miniband is not a single scale. It conflates three physically distinct widths, namely the intrinsic Jackiw--Rebbi hybridization, the zone-edge upturn, and the bias-induced mid-zone bump, which follow different scaling laws and are treated separately in Sec.~\ref{sec:scaling}. We also discuss a single-orbital bond-order realization and several experimental platforms.

		The paper is organized as follows. Section~\ref{sec:model} introduces
		the model, the folded two-band Hamiltonian, and the fixed-$k_y$ Dirac
		reduction, with attention to where the linearization breaks down.
		Section~\ref{sec:jr} gives the Jackiw--Rebbi solution for an
		isolated wall. Section~\ref{sec:ssh} develops the effective SSH
		description and the bias-induced dimerization. Section~\ref{sec:helical}
		derives the boundary phase and the helical node condition, including
		the lattice parity rule for the node positions.
		Section~\ref{sec:scaling} defines and derives the three widths of the
		central miniband, $W_{\rm JR}$, $W_{\rm edge}$, and $W_{\rm bump}$, and
		treats sharp textures. Section~\ref{sec:lattice} discusses the valley
		doubling of the model and a single-orbital bond-order realization.
		Section~\ref{sec:numerics}
		presents exact-diagonalization tests of every mechanism.
		Section~\ref{sec:experiment} collects experimental proposals.
		Section~\ref{sec:discussion} discusses robustness, interaction scales, and
		open directions, and Section~\ref{sec:summary} summarizes.
		
		\section{Model and folded Hamiltonian}
		\label{sec:model}
		
We consider a general two-band model on the square lattice with two
components per unit cell, collected in the two-component field $c_i=(a_i,b_i)^T$:
\begin{equation}
	H=-t\sum_{\langle ij\rangle}\big(c_i^\dagger\sigma_3\,c_j+\mathrm{H.c.}\big)
	+\sum_{i}m(x_i)\,c_i^\dagger\sigma_2\,c_i,
	\label{eq:h0}
\end{equation}
where the Pauli matrices act on an orbital or sublattice degree of
freedom, and $m(x)$ is an inter-orbital (or inter-sublattice)
hybridization whose amplitude varies slowly along $x$,
\begin{equation}
	m(x)=m+m_0\cos(Qx),\qquad Q=\frac{2\pi}{L}.
	\label{eq:mx}
\end{equation}
Here $m_0$ is the modulation amplitude, $m$ is a uniform bias, and $L$
is the modulation period, commensurate with the lattice and ranging up
to of order $10^3$ sites.

In the orbital reading, the opposite sign of the $\sigma_3$ hopping is
an orbital polarization of the intra-orbital dispersions,
$\epsilon_a(\mathbf k)=\epsilon(\mathbf
k)=-2t(\cos k_x+\cos k_y)$ and $\epsilon_b(\mathbf
k)=-\epsilon(\mathbf k)$, and the Hamiltonian acts on
$\Psi_{\mathbf k}=(a_{\mathbf k},b_{\mathbf k})^T$ as
\begin{equation}
	H(\mathbf k)=\epsilon(\mathbf k)\,\sigma_3+m(x)\,\sigma_2.
	\label{eq:hfold}
\end{equation}
Because the hybridization is on-site it carries no momentum
dependence, and Eq.~\eqref{eq:hfold} is exact for any texture $m(x)$.

The sublattice reading is realized by a single-orbital system with a
$(\pi,\pi)$ charge bond order (CBO), a modulation of the $x$-oriented
bonds that folds the Brillouin zone: the two components are the
valleys $\mathbf k$ and $\mathbf k+\mathbf Q$, and the intra-component
$\sigma_3$ term arises naturally because the square-lattice dispersion
obeys the exact nesting condition
\begin{equation}
	\epsilon(\mathbf k+\mathbf Q)=-\epsilon(\mathbf k),
	\qquad \mathbf Q=(\pi,\pi),
	\label{eq:nesting}
\end{equation}
so that the identification
\begin{equation}
	c_a(\mathbf k)\equiv c(\mathbf k),\qquad
	c_b(\mathbf k)\equiv c(\mathbf k+\mathbf Q)
	\label{eq:map}
\end{equation}
brings the folded Hamiltonian to the form \eqref{eq:hfold}, with the
bond modulation supplying the mass. The folded mass then carries a
bond form factor, innocuous near the band crossings and vanishing at
the van Hove points; its precise form, and the resulting differences
from the on-site realization, are worked out in
Sec.~\ref{sec:lattice}. All analytics below use Eq.~\eqref{eq:hfold},
and all numerics diagonalize the model \eqref{eq:h0} directly
(Sec.~\ref{sec:numerics}).
		
		\subsection{Fixed-$k_y$ Dirac reduction}
		
		Because $m(x_i)$ varies only along $x$, $k_y$ is conserved. At fixed
		$k_y$ the two folded bands cross on the degeneracy line
		\begin{equation}
			\cos k_x^0=-\cos k_y,\qquad k_x^0=\pi-k_y\ \ (k_y\in[0,\pi]),
			\label{eq:line}
		\end{equation}
		with the true cone velocity along it denoted $u_x(k_y)=2t\sin k_y$. For the
		Jackiw--Rebbi construction of Sec.~\ref{sec:jr} and the boundary phase
		of Sec.~\ref{sec:helical} it is convenient to expand instead about the
		fixed symmetric point $k_x=\pi/2$ of the folded Fermi line,
		$\epsilon(\pi/2+q,k_y)=2t\,q-2t\cos k_y+O(q^2)$, which tracks the
		degeneracy line through $q=\cos k_y$ near $k_y=\pi/2$. This yields
		the $1+1$ dimensional Dirac Hamiltonian
		\begin{equation}
			H_D(k_y)=\big[-iv_x\partial_x+\mu(k_y)\big]\sigma_3+m(x)\,\sigma_2,
			\label{eq:hd}
		\end{equation}
		with a $k_y$-independent velocity and a scalar detuning
		\begin{equation}
			v_x=2t,\qquad \mu(k_y)=-2t\cos k_y.
			\label{eq:lin}
		\end{equation}
		At the symmetric point $k_y=\pi/2$ the detuning vanishes and the
		expansion point coincides with the degeneracy line, $k_x^0=\pi/2$.
		The two velocities should not be conflated: the constant $v_x=2t$
		enters the chiral structure and the gauge phase of
		Sec.~\ref{sec:helical}, while the tunneling decay constants between
		walls are set by the true velocity on the degeneracy line, denoted
		$u_x(k_y)=2t\sin k_y$ in what follows [Sec.~\ref{sec:ssh}].
		
		Two properties of Eq.~\eqref{eq:hd} organize everything that follows.
		First, at $\mu=0$ the Hamiltonian anticommutes with $\sigma_1$,
		\begin{equation}
			\{H_D(k_y=\pi/2),\,\sigma_1\}=0,
			\label{eq:chiral}
		\end{equation}
		so $\sigma_1$ is the chiral operator of the fixed-$k_y$ Dirac problem.
		With the mass on $\sigma_2$ and the kinetic term on $\sigma_3$, only
		$\sigma_1$ anticommutes with both; this choice of chiral generator fixes
		the internal polarization of the zero modes constructed in
		Sec.~\ref{sec:jr}. Away from $k_y=\pi/2$ the scalar term
		$\mu(k_y)\sigma_3$ also anticommutes with $\sigma_1$: it shifts the
		Dirac point in energy without hybridizing the two valleys, and the
		zero-mode construction below is unaffected because $\mu$ enters the
		zero-energy condition only as a phase (Sec.~\ref{sec:jr}).
		
		Second, the reduction has two controlled limitations. (i) Near
		$k_y=0,\pi$ the degeneracy line approaches the van Hove saddle of the
		square lattice and the true cone velocity $2t\sin k_y\to0$: the
		linearization fails in a window around these lines, and the central miniband merges there into the bulk
		valley band edges, producing the zone-edge upturn analyzed in
		Sec.~\ref{sec:wedge}.  Statements made below about flat bands, helical
		nodes, and dimerization apply to the broad region around
		$\pm\pi/2$ where $u_x(k_y)$ is finite; in particular the flat zero-mode band
		of the strong-modulation regime is flat in the $k_y$-region around
		$\pm\pi/2$. (ii) The two-valley description assumes
		that no other Fourier component of the texture reconnects $c_{\mathbf
			k}$ and $c_{\mathbf k+\mathbf Q}$; this holds exactly for any
		single-harmonic modulation $m(x)$, and multi-harmonic textures
		reconnect the cones (Sec.~\ref{sec:lattice}).
		
		Finally, the lattice Hamiltonian Eq.~\eqref{eq:h0} anticommutes with
		the orbital-interchange operation $\Gamma:\,a_j\leftrightarrow b_j$:
		the intra-orbital hoppings change sign owing to their opposite signs,
		and the on-site hybridization changes sign under the interchange (the
		$\sigma_2$ form). The spectrum is therefore exactly particle-hole
		symmetric at every $(k_y,k_x^{\rm sc})$, and the lowest positive
		miniband is nonnegative, touching zero only at isolated nodes. In the
		folded basis $\Gamma$ is precisely the $\sigma_1$ of
		Eq.~\eqref{eq:chiral}, here exact on the lattice for any texture.
		
		\subsection{Symmetry protection and topological content}
		\label{sec:topology}
		
		It is worth stating precisely in what sense the model is topological.
		The zero modes are topological bound states in the one-dimensional
		Jackiw--Rebbi sense: each simple zero of $m(x)$ is a topological defect
		of the mass field \cite{teo2010}, the number of zero modes equals the
		number of mass sign changes, and their pinning to zero energy is
		enforced by symmetry
		rather than by energetics \cite{ryu2002,chiu2016}.  On the lattice the protecting symmetry is
		the orbital-interchange operation $\Gamma$ introduced above, which is
		exact for any on-site hybridization texture and is broken only by
		further-neighbor intra-orbital hopping or by additional inter-orbital
		matrix structures (such as a winding phase of the texture); in the
		continuum it is the $\sigma_1$ chiral symmetry Eq. \eqref{eq:chiral} at
		$k_y=\pi/2$, together with the fact that $\mu(k_y)$ enters the
		zero-mode condition only as a phase (Sec.~\ref{sec:jr}).  The dimerized
		SSH ring of Sec.~\ref{sec:ssh} adds a genuine $\mathbb{Z}$ winding
		number [Eq.~\eqref{eq:wind}], $\nu=1$ for $m>0$ versus $\nu=0$ for
		$m<0$, and the helical nodes of Sec.~\ref{sec:helical} are enforced by
		the boundary phase.
		By contrast, the model carries no two-dimensional topological
		invariant: the folded Hamiltonian is real, the Chern number vanishes,
		and the flat band is not a topological flat band in the
		quantum-geometry sense.  In particular the flatness itself is not
		topologically protected: the zero-mode energies are pinned by symmetry,
		but the bandwidth $W_{\rm JR}$ is set by wall-to-wall tunneling and is
		exponentially small rather than zero at any finite $L$
		(Sec.~\ref{sec:scaling}).
		
		\section{Jackiw--Rebbi modes of an isolated mass wall}
		\label{sec:jr}
		
		For a mass wall centered at $x_n$, $m(x_n)=0$, we expand
		$m(x)\simeq \kappa_n (x-x_n)$ with local slope $\kappa_n=m'(x_n)$, and
		suppress $\mu(k_y)$, which is a pure spectator for the zero-energy
		condition.  Acting with $H_D(k_y)\chi(x)=0$, where
		\begin{equation}
			H_D(k_y)=(-iv_x\partial_x)\sigma_3+\kappa_n(x-x_n)\sigma_2,
			\label{eq:hdwall}
		\end{equation}
		the two components decouple once $\chi$ is chosen as an eigenstate of the
		chiral operator $\sigma_1$, $\sigma_1\chi_c=c\,\chi_c$, $c=\pm1$.  Using
		$\sigma_3\chi_c=\chi_{-c}$ and $\sigma_2\chi_c=-ic\chi_{-c}$, the equation
		reduces to the scalar condition
		\begin{equation}
			\big[v_x\partial_x+c\,\kappa_n(x-x_n)\big]f(x)=0,
			\label{eq:scalar}
		\end{equation}
		with normalizable solution
		\begin{align}
			\psi_{n,c}(x)&=\frac{1}{\sqrt{\mathcal N}}\,
			\exp\!\left[-\frac{c\,\kappa_n}{2v_x}(x-x_n)^2\right]\chi_c,
			\nonumber\\
			c&=\mathrm{sgn}(\kappa_n).
			\label{eq:jr}
		\end{align}
		Only the branch $c=\mathrm{sgn}(\kappa_n)$ decays: a wall with
		$\kappa_n>0$ binds a $\sigma_1=+1$ mode, and an anti-wall with
		$\kappa_n<0$ binds a $\sigma_1=-1$ mode.  Because the cosine texture
		alternates wall and anti-wall, the two zero modes of a period carry
		opposite $\sigma_1$ eigenvalues.  This alternating internal
		polarization is the continuum counterpart of the sublattice polarization
		of domain-wall states in polyacetylene, and it is what allows the two
		walls to be viewed as the two sites of an effective SSH unit cell
		(Sec.~\ref{sec:ssh}).  The localization length at the expansion point is
		\begin{equation}
			\xi_n=\sqrt{\frac{v_x}{|\kappa_n|}},
			\qquad
			|\kappa_n|=\frac{2\pi}{L}\sqrt{m_0^2-m^2}.
			\label{eq:xi}
		\end{equation}
		Note that $|\kappa_n|\propto 1/L$ for a smooth texture: walls become
		parametrically softer as the period grows, a point that controls the
		bandwidth scaling of Sec.~\ref{sec:scaling}.
		
		The argument generalizes verbatim to any texture $m(x)$ with simple,
		well-separated zeros.  Two ingredients are essential and will be assumed
		throughout: (i) the two valleys $c_{\mathbf k}$ and $c_{\mathbf k+Q}$ that
		the mass term couples must not be reconnected by any other Fourier
		component of the texture (exact chiral pinning), and (ii) the Fermi
		velocity $u_x(k_y)=2t\sin k_y$ must remain finite, which excludes a
		van Hove neighborhood of $k_y=0,\pi$ whose width in $k_y$ scales as
		$m_0/t$ and vanishes with the modulation strength.
		
\begin{figure}[t]
			\centering
			\includegraphics[width=\columnwidth]{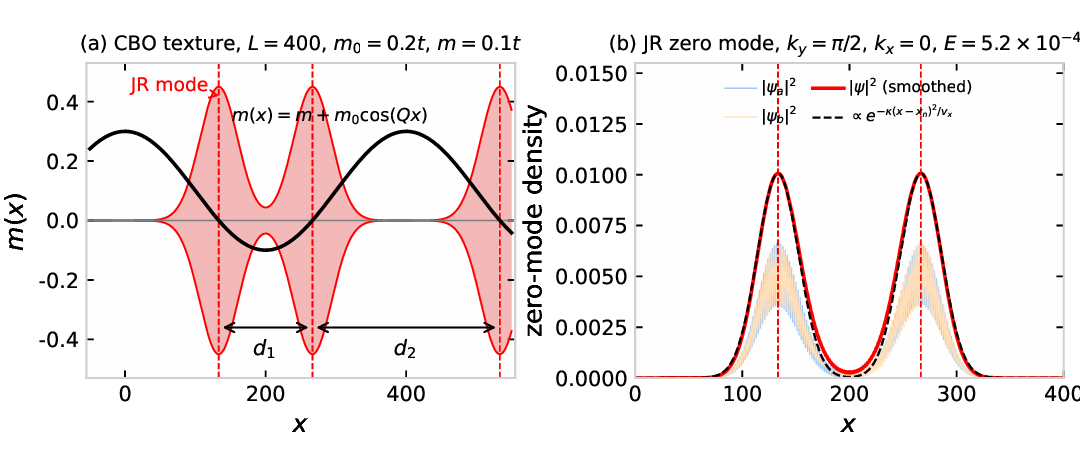}
			\caption{(a) Mass texture $m(x)=m+m_0\cos(2\pi x/L)$ (here $m=0.1t$,
				$m_0=0.2t$, $L=400$), its two zeros per period (dashed lines), the
				Jackiw--Rebbi envelopes localized at the zeros (shaded), and the two
				barrier widths $d_1\neq d_2$ that encode the bias-induced dimerization
				of the effective SSH ring.  (b) Zero-mode density at $k_y=\pi/2$,
				$k_x^{\rm sc}=0$ from exact diagonalization ($|\psi_a|^2$,
				$|\psi_b|^2$, and the smoothed total), compared with the analytic
				envelope of Eq.~\eqref{eq:jr} (dashed).  The hybridized mode sits at
				$E=5.2\times10^{-4}t$, the bump scale of Sec.~\ref{sec:wbump}.}
			\label{fig:schematic}
		\end{figure}

		Figure~\ref{fig:schematic} illustrates the construction on the lattice:
		a biased texture with two zeros per period, the Jackiw--Rebbi envelopes
		localized at the walls, and the unequal barrier widths $d_1\neq d_2$
		that will encode the dimerization of Sec.~\ref{sec:ssh}.
		
		\section{Effective SSH description and bias-induced dimerization}
		\label{sec:ssh}
		
		\subsection{Tunneling network at fixed $k_y$}
		
		Let $t_1(k_y)$ and $t_2(k_y)$ be the tunneling amplitudes through the two
		mass barriers separating the two walls of a supercell.  Retaining only
		nearest-wall tunneling, the low-energy Hamiltonian in the zero-mode basis
		$\{|1\rangle,|2\rangle\}$ is the two-site SSH problem
		\begin{align}
			H_{\rm eff}(k_y;k_x^{\rm sc})&=
			\begin{pmatrix}
				0 & t_1-t_2\,e^{i\Phi}\\[2pt]
				t_1-t_2\,e^{-i\Phi} & 0
			\end{pmatrix},
			\label{eq:ssh}\\
			E_\pm&=\pm\big|t_1-t_2e^{i\Phi}\big|.
			\label{eq:sshe}
		\end{align}
		Here $k_x^{\rm sc}$ is the supercell crystal momentum and $\Phi$ is the
		phase accumulated around the ring, derived in Sec.~\ref{sec:helical}.
		The relative minus sign between the two tunneling paths reflects the
		opposite $\sigma_1$ polarization of the two walls
		(Sec.~\ref{sec:jr}); equivalently it can be absorbed into a $\pi$ shift
		of the origin of $\Phi$, and we choose the convention in which the
		zero-bias nodes occur at $\Phi=0$.
		Within WKB, the barrier actions
		$S_{1,2}(k_y)=u_x(k_y)^{-1}\int_{\rm barrier}|m(x)|\,dx$, with
		$u_x(k_y)=2t\sin k_y$ the true cone velocity on the degeneracy line
		Eq. \eqref{eq:line}, evaluate to
		\begin{align}
			S_{1}&=\frac{2\big[m_0\sin\theta_0+m\,\theta_0\big]}{Qu_x(k_y)},
			\nonumber\\
			S_{2}&=\frac{2\big[m_0\sin\theta_0-m(\pi-\theta_0)\big]}{Qu_x(k_y)},
			\label{eq:actions}
		\end{align}
		with $\theta_0=\arccos(-m/m_0)\in[\pi/2,\pi]$ for $0\leq m<m_0$, so that
		$t_{1,2}\propto e^{-S_{1,2}}$.  Here $S_1$ is the action of the wide
		barrier, where $m(x)>0$, and $S_2$ that of the narrow barrier, where
		$m(x)<0$; at $m=0$ they reduce to $S_1=S_2=m_0L/\pi u_x(k_y)$. These actions
		grow linearly with the period, so the central band flattens
		exponentially as $L$ grows [Fig.~\ref{fig:bands}(a)]. Both the
		helical oscillation height at $m=0$ and the bias-induced bump height
		are controlled by these actions: the measured decay rates in $L$ agree
		with $S_{1,2}/L$ to within about five percent
		(Secs.~\ref{sec:scaling} and \ref{sec:numerics}).  Two exact structural
		consequences follow.
		
		First, at zero bias the two barriers are related by a translation of
		$L/2$ combined with $m_0\to-m_0$, and under this operation
		$t_1\leftrightarrow t_2$; hence the effective SSH ring sits exactly at the
		dimerization transition, $|t_1|=|t_2|$.  The resulting low-energy bands
		touch, and since they touch with opposite chirality polarization the
		touching is the source of the helical minibands of
		Sec.~\ref{sec:helical}.
		
		Second, the action difference accumulated by a uniform bias $m$ is
		\begin{equation}
			S_1-S_2=\frac{2\pi m}{Q\,u_x(k_y)}=\frac{mL}{u_x(k_y)},
			\qquad
			\frac{t_1}{t_2}=e^{\,-mL/u_x(k_y)},
			\label{eq:dimer}
		\end{equation}
		i.e., for $m>0$ the thinned narrow barrier is the more transparent one,
		$t_2>t_1$, and the effective SSH ring is dimerized by an amount
		controlled
		exponentially by the total flux-like quantity $mL/u_x(k_y)$
		(with $u_x(\pi/2)=v_x$ at mid-zone).  The
		associated winding number is that of the off-diagonal element of
		Eq.~\eqref{eq:ssh},
		\begin{equation}
			\nu=\frac{1}{2\pi}\int_0^{2\pi}\!d\Phi\,\partial_\Phi\arg q(\Phi),
			\qquad q(\Phi)=t_1-t_2\,e^{-i\Phi}.
			\label{eq:wind}
		\end{equation}
		The loop $q(\Phi)$ is a circle of radius $|t_2|$ centered at $t_1$,
		which encloses the origin if and only if $|t_2|>|t_1|$; by
		Eq.~\eqref{eq:dimer} this gives $\nu=1$ for $m>0$ and $\nu=0$ for
		$m<0$, so the sign of the bias is a topological invariant of the
		hybridized ring.  The integral is taken at fixed $k_y$: since
		$d\Phi/dk_x^{\rm sc}=L$ exactly, $\Phi$ advances by precisely $2\pi$
		as $k_x^{\rm sc}$ traverses the supercell Brillouin zone at any $k_y$,
		and the $\cos k_y$ offset merely shifts the starting point of the loop.
		The $k_y$ dependence of $t_{1,2}$ changes the loop geometry but never
		its enclosure of the origin, because $u_x(k_y)=2t\sin k_y>0$ on
		$(0,\pi)$ keeps the ordering of $|t_1|,|t_2|$ uniform.  Hence $\nu$ is
		independent of $k_y$ wherever the two-band reduction holds; it can
		change only where the ring gap closes ($m=0$), and it is undefined at
		the van Hove lines $k_y=0,\pi$, where the description breaks down.  The miniband splitting and bandwidth
		that result are
		\begin{equation}
			\Delta_{\rm DW}=2\big||t_1|-|t_2|\big|,
			\qquad
			W=2\min(|t_1|,|t_2|).
			\label{eq:gapw}
		\end{equation}
		Equation~\eqref{eq:dimer} is the continuum statement of the
		pairing picture: the two zeros of the period approach each other under
		bias, the barrier between them thins, and the flat band converts into a
		pair of dispersive, particle-hole symmetric bands.  Because both
		actions [Eq. \eqref{eq:actions}] grow linearly with $L$, every bias-induced
		feature of the central miniband is exponential in the period: the
		mid-zone dispersion opened by the bias (the bump of
		Sec.~\ref{sec:scaling}) is a mesoscopic effect, and the flat band
		becomes robust against bias as $L\to\infty$.  The quantitative
		verification of the actions, and the sense in which the amplitudes
		$t_{1,2}$ follow or depart from simple exponentials of
		Eq.~\eqref{eq:actions}, is developed in Sec.~\ref{sec:scaling}.
		
		\subsection{Where the dimerization gap opens}
		
		The actions [Eq. \eqref{eq:actions}] diverge as $u_x(k_y)=2t\sin k_y\to0$,
		so the hybridization is strongly $k_y$ dependent: it is maximal near
		$k_y=\pm\pi/2$ where the velocity is maximal, and is suppressed near the
		van Hove lines $k_y=0,\pi$.  This velocity dependence is borne out
		directly by the helical envelope of the unbiased lattice bands: the
		measured oscillation amplitude at $k_y=0.75$ is $0.20$ of its
		$k_y=\pi/2$ value, against the WKB prediction
		$e^{-S_0(1/\sin k_y-1)}=0.23$ with $S_0=m_0L/\pi v_x$
		(Sec.~\ref{sec:numerics}).  Consequently the bias-induced gap
		$\Delta_{\rm DW}$ opens first and essentially uniformly around
		$k_y=\pm\pi/2$, while the bands remain nearly flat and gapless-looking
		close to $k_y=0,\pi$ until the bias exceeds a larger threshold. This is
		borne out by the numerics of Sec.~\ref{sec:numerics}, and it means that
		``the bias gaps the flat band'' is precise only in the window where the
		Dirac linearization with finite velocity applies.
		
		\section{Helical minibands from the boundary phase}
		\label{sec:helical}
		
		\begin{figure*}[t]
			\centering
			\includegraphics[width=0.9\textwidth]{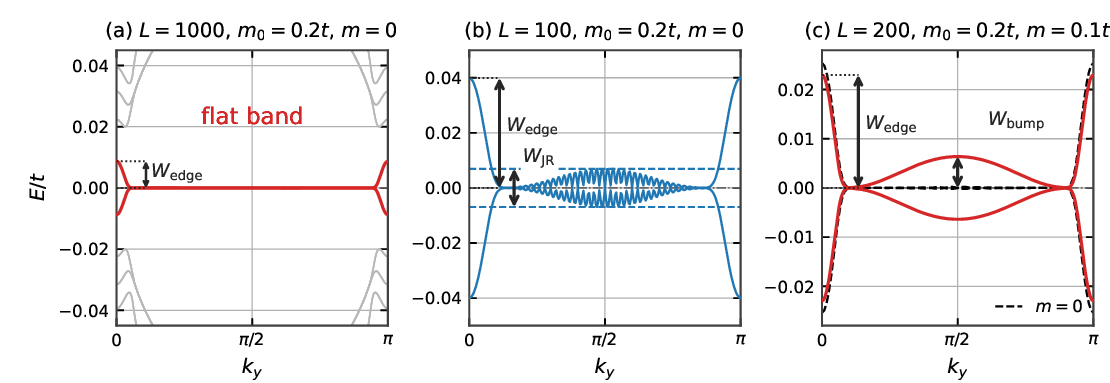}
			\caption{Low-energy spectrum of Eq.~\eqref{eq:hfold} versus $k_y$
				($k_x^{\rm sc}=0$): the three regimes and the three widths of the
				central miniband.
				(a) Flat-band regime ($L=1000$, $m_0=0.2t$, $m=0$): the central band
				(red) is pinned at zero; gray curves are the Dirac ladder with
				$E_1\simeq\sqrt{8\pi m_0/L}\,t=0.070t$.
				(b) Helical regime ($L=100$, $m_0=0.2t$, $m=0$): the central band
				(blue) oscillates through nodes at $k_y=\pi/2+2\pi n/L$ (the $n=0$
				node at $\pi/2$ is exact for $L\equiv0\ (\mathrm{mod}\ 4)$); dashed
				lines mark $\pm W_{\rm JR}/2$.
				(c) Biased central band ($L=200$, $m_0=0.2t$, $m=0.1t$, red) against
				the $m=0$ reference (dashed); the arrow at $k_y=\pi/2$ marks the bump
				height $W_{\rm bump}$.
				In all panels the arrows near $k_y=0$ mark the zone-edge height
				$W_{\rm edge}$ measured from $E=0$; at $m=0$ this coincides with the
				max-min definition [Eq. \eqref{eq:defedge}], because the miniband minimum
				vanishes [in (b) the zone corner $k_y=0$ itself lies on the node
				lattice].}
			\label{fig:bands}
		\end{figure*}
		
		We now explain the mechanism behind the helical regime, which is the
		regime of small $L$ or weak modulation where $t_1,t_2$ are not small and
		the low-energy physics is a coherent ring rather than isolated walls
		[Fig.~\ref{fig:bands}(b)].
		
		Removing the scalar $\mu(k_y)=-2t\cos k_y$ from $H_D$ requires the gauge
		transformation $U(x)=\exp[i\mu(k_y)x/v_x]$, since
		$U^\dagger(\mu\sigma_3)U$ combined with the derivative shift eliminates
		$\mu$.  Under supercell periodic boundary conditions,
		$\psi(x+L)=e^{ik_x^{\rm sc}L}\psi(x)$, the transformed spinor obeys
		twisted boundary conditions with twist $\mu L/v_x$, which is generically
		not a multiple of $2\pi$: the gauge transformation is not single-valued
		on the ring.  The twist is absorbed as a shift of the supercell momentum,
		so that the phase winding around the SSH ring in Eq.~\eqref{eq:ssh} is
		\begin{equation}
			\Phi(k_y,k_x^{\rm sc})
			=\Big(k_x^{\rm sc}-\frac{\mu(k_y)}{v_x}\Big)L
			=\big[k_x^{\rm sc}+\cos k_y\big]L,
			\label{eq:phase}
		\end{equation}
		where we used $\mu/v_x=-\cos k_y$, the ratio of the fixed-frame values
		[Eq. \eqref{eq:lin}] of the expansion about $k_x=\pi/2$.
		At $m=0$, where $|t_1|=|t_2|\equiv t_c$, the minibands are
		\begin{equation}
			E_\pm(k_y,k_x^{\rm sc})
			=\pm 2t_c(k_y)\,\big|\sin\tfrac{\Phi}{2}\big|,
			\label{eq:helix}
		\end{equation}
		with protected nodes whenever $\Phi=0\ (\mathrm{mod}\ 2\pi)$, i.e., along
		the lines
		\begin{equation}
			L\cos k_y=2\pi n,\qquad n\in\mathbb Z,
			\label{eq:nodes}
		\end{equation}
		at $k_x^{\rm sc}=0$ (and symmetry-related points).  For $L\gtrsim2\pi$
		the BZ therefore hosts $\sim 2L/\pi$ nodal lines in $k_y$, evenly spaced
		by $\Delta k_y=2\pi/L$ around $\pi/2$: a ladder of helical crossings.

		Equation \eqref{eq:nodes} is the $\pi/2$-linearized form of the lattice
		condition: the true cone sits at $k_x^0=\pi-k_y$ [Eq.~\eqref{eq:line}],
		so cone-on-grid commensurability reads $L(\pi/2-k_y)=2\pi n$ (for
		$L\equiv0\ (\mathrm{mod}\ 4)$), of which Eq.~\eqref{eq:nodes} is the
		leading order since $\cos k_y=\pi/2-k_y+O[(k_y-\pi/2)^3]$.  The
		measured node positions at $L=100$ track the lattice form to within a
		few $10^{-3}$, a few percent of the spacing $2\pi/L$, out to the tenth
		node and drift only toward the van Hove lines, whereas the fixed-frame
		form [Eq. \eqref{eq:nodes}] errs by $(k_y-\pi/2)^3/6$, about $0.05$ at the
		tenth node.

		On the lattice we find one refinement of Eq.~\eqref{eq:nodes}: the
		oscillation phase alternates with the parity of $L/2$.  For $L\equiv0\
		(\mathrm{mod}\ 4)$ the nodes sit at $k_y=\pi/2+2\pi n/L$, including the
		$n=0$ member at $k_y=\pi/2$; for $L\equiv2\ (\mathrm{mod}\ 4)$ the
		pattern shifts by half a spacing, $k_y=\pi/2+(2n+1)\pi/L$, and the
		miniband has an antinode at $k_y=\pi/2$ instead.  The origin is the
		sign $(-1)^{L/2}$ carried by the wall-to-wall tunneling amplitude
		through a barrier of $L/2$ sites, which shifts the interference
		condition [Eq. \eqref{eq:nodes}] by $\pi$.  The point $k_y=\pi/2$ remains
		special throughout: there $\mu=0$ identically and the two-band lattice
		Hamiltonian anticommutes with $\sigma_1$ as a matrix, so the
		spectrum is exactly particle-hole symmetric at $k_y=\pi/2$ for any
		texture profile; whether a member of the ladder sits exactly at zero
		there is decided by the parity rule above.  The exact nodes approach
		the lines [Eq. \eqref{eq:nodes}] as $L$ grows at fixed $m_0$: deviations of
		order ten percent at $L\sim16$ shrink to about one percent at $L=64$
		and below $10^{-4}$ at $L\sim10^2$, while the minima of the wiggling
		miniband at fixed $k_x^{\rm sc}=0$ remain evenly spaced by $2\pi/L$
		around $\pi/2$ throughout (Sec.~\ref{sec:numerics}).

		The physical picture is that of a flux-threaded ring: $k_y$ tunes an
		effective flux $\mu L/v_x$ through the supercell, and each time the flux
		passes an integer value the ring spectrum becomes gapless.  The
		``helical'' character follows because the two crossing branches carry
		opposite $\sigma_1$ polarization and disperse in opposite directions.
		This mechanism differs from both the dispersive chiral modes of mass
		superlattices \cite{demartino2023,anand2025}, which are unidirectional
		and live on a single wall, and from accidental band touchings: here the
		nodes are enforced by the boundary phase [Eq. \eqref{eq:phase}] and are
		therefore robust against any deformation of $m(x)$ that preserves its
		simple zeros and the two-valley structure.

		\section{Three widths of the central miniband}
		\label{sec:scaling}
		
		The central miniband $E_0(k_y,k_x^{\rm sc})$ carries three distinct
		width-like quantities, with three different mechanisms and three
		different scaling laws.  We define them operationally:
		\begin{align}
			&W_{\rm JR}=2\max_{n}\big|E_0\big(\tfrac{\pi}{2}+\tfrac{n\pi}{L},0\big)\big|,
			\label{eq:defjr}\\
			&W_{\rm edge}=\max_{k_y}E_0(k_y,0)-\min_{k_y}E_0(k_y,0),
			\label{eq:defedge}\\
			&W_{\rm bump}(m)=\big|E_0\big(\tfrac{\pi}{2},0\big)\big|_{m}.
			\label{eq:defbump}
		\end{align}
		These are, respectively, the helical height at $m=0$, the zone-edge
		upturn, and the bias bump [Fig.~\ref{fig:bands}(c)].
		In Eq.~\eqref{eq:defjr} the maximum runs over a few integers
		$|n|\leq2$, which covers both parities of the node rule of
		Sec.~\ref{sec:helical}.  We stress that the total max-min bandwidth of
		the miniband equals $W_{\rm edge}$ at all but the smallest periods:
		the three quantities must not be conflated, and we discuss them in
		turn.  All numerical statements below refer to exact diagonalization of
		the model Hamiltonian [Eqs.~\eqref{eq:h0}--\eqref{eq:hfold}] on
		supercells up to $L=10^3$ with $m_0=0.2t\ldots2t$
		(Sec.~\ref{sec:numerics}).
		
		\subsection{Helical height: the WKB exponential restored}
		\label{sec:wjr}
		
\begin{figure}[t]
			\centering
			\includegraphics[width=\columnwidth]{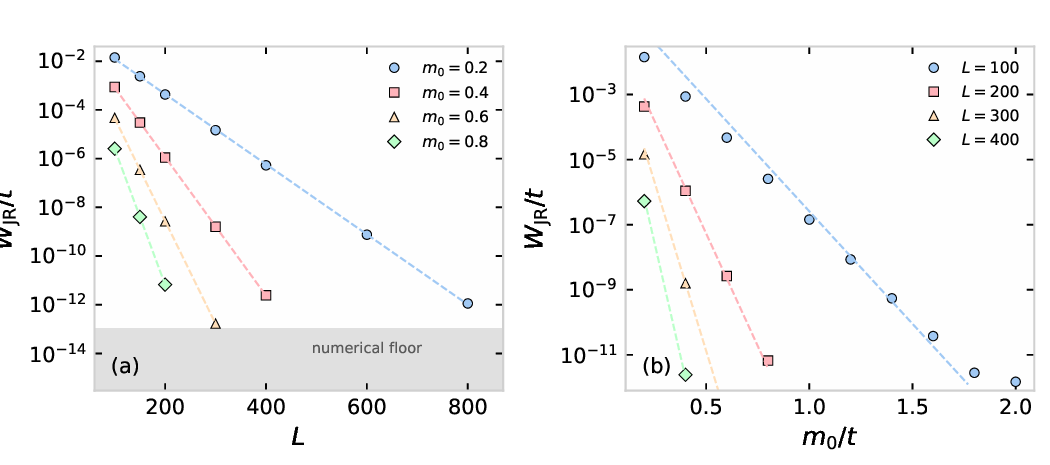}
			\caption{(a) Semilog plot of the helical height $W_{\rm JR}$ versus $L$
				at $m=0$ for $m_0=0.2t\ldots0.8t$ (symbols), with exponential fits
				$C(m_0)e^{-m_0L/\pi v_x}$ (dashed): measured decay rates $0.0331$,
				$0.0656$, $0.0972$, and $0.1286$ per site against the barrier actions
				$m_0/\pi v_x=0.0318$, $0.0637$, $0.0955$, and $0.1273$.  The gray band
				marks the $\sim10^{-13}t$ solver noise floor.  (b) $W_{\rm JR}$ versus
				$m_0$ at fixed $L=100\ldots400$; dashed lines are guides with the fixed
				theoretical slope $-L/\pi v_x$.}
			\label{fig:wjr}
		\end{figure}

		At zero bias the oscillation amplitude of the helical miniband between
		its nodes measures the wall-to-wall tunneling $t_c$ through either
		barrier, $W_{\rm JR}=4|t_c|$ within the SSH reduction
		[Eq. \eqref{eq:helix}].  The barrier action at $m=0$ is
		\begin{equation}
			S_0=\frac{1}{v_x}\int_0^{L/2}|m(x)|\,dx
			=\frac{m_0L}{\pi v_x},
			\label{eq:s0}
		\end{equation}
		so the helical height is expected to follow
		\begin{equation}
			W_{\rm JR}\simeq C(m_0)\,e^{-m_0L/\pi v_x}.
			\label{eq:wjr}
		\end{equation}
		Equation~\eqref{eq:wjr} is confirmed quantitatively
		(Fig.~\ref{fig:wjr}): on the grid $L=100\ldots1000$ the measured decay
		rates are $0.0331$, $0.0656$, $0.0972$, and $0.1286$ per site at
		$m_0=0.2t,0.4t,0.6t,0.8t$, against $m_0/\pi v_x=0.0318$, $0.0637$,
		$0.0955$, and $0.1273$ (with $v_x=2t$), a one-to-four percent deviation
		attributable to lattice corrections of the velocity
		and of $\langle|m|\rangle$.  The prefactor is of order the modulation
		scale itself, $C(0.2t)\simeq0.34t$ and $C(0.8t)\simeq0.97t$.  Note the
		contrast with the naive worry that soft walls, $\kappa\propto1/L$ in
		Eq.~\eqref{eq:xi}, might destroy the exponential: the Gaussian zero
		modes widen as $\sqrt L$, but the barrier they tunnel through widens as
		$L$, and the action [Eq. \eqref{eq:s0}] wins.  Because $W_{\rm JR}$ falls
		below the double-precision noise floor $\sim10^{-13}t$ once
		$m_0L/\pi v_x\gtrsim28$, the measurable window shrinks with $m_0$
		(for $m_0=0.5t$ one needs $L\lesssim370$; for $m_0=2t$ only
		$L\lesssim100$): the exponential law itself asserts that the helical
		height of a large-period sample is exactly zero for all practical
		purposes [Fig.~\ref{fig:bands}(a)].
		
		\subsection{Zone-edge upturn: merging with the bulk band edges}
		\label{sec:wedge}
		
		The upturn of the miniband near $k_y=0,\pi$ has a different origin.
		The transverse momentum enters the folded Hamiltonian [Eq. \eqref{eq:hfold}]
		only through the valley detuning $\mu(k_y)\sigma_3=-2t\cos
		k_y\,\sigma_3$, which pushes the two folded valleys apart in opposite
		directions.  Around $k_y=\pi/2$ the detuning vanishes and the zero-mode
		state is valley balanced, $\langle\sigma_3\rangle\simeq0$.  Toward the
		zone edge the state repolarizes continuously: at $L=200$, $m_0=0.2t$
		and $k_y\to0$ its valley weights approach $\sin^2(\pi/8)$ and
		$\cos^2(\pi/8)$, while its real-space support shrinks from both walls
		to a single wall (the participation ratio drops from $\simeq93$ to
		$\simeq23$ sites).  At the same time the whole Dirac ladder compresses
		toward zero (the first rung moves from $0.151t$ at $k_y=\pi/2$ to
		$0.065t$ at $k_y=0$ in the same system), because at $k_y=0,\pi$ the two
		valley band edges $\mp2t(\cos k_x\pm1)$ both touch $E=0$ quadratically
		at $k_x=\pi$: the miniband merges into the bulk continuum, and the
		upturn is the finite-size boundary layer of this merging.
		
\begin{figure}[t]
			\centering
			\includegraphics[width=0.9\columnwidth]{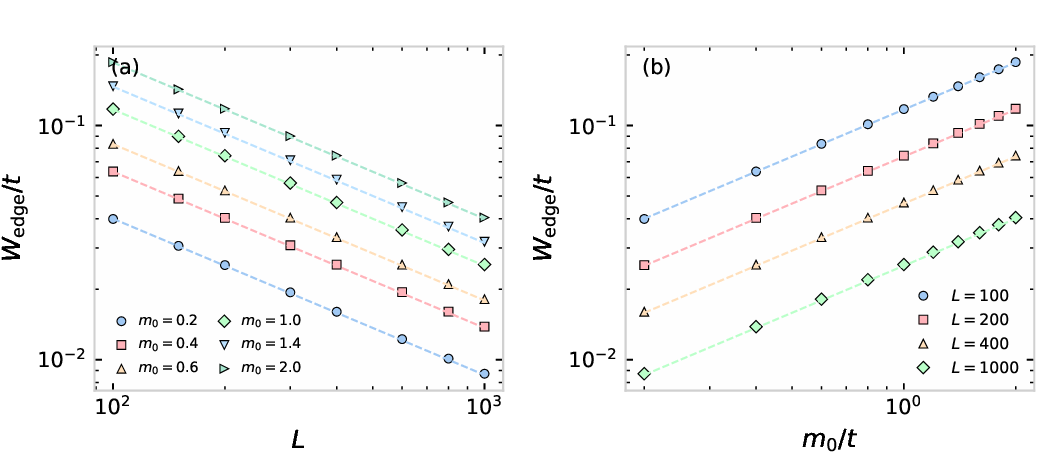}
			\caption{(a) Log-log plot of the zone-edge upturn $W_{\rm edge}$
				[Eq.~\eqref{eq:defedge}] versus $L$ for $m_0=0.2t\ldots2t$ (symbols),
				with the law $0.74\,[t(m_0Q)^2]^{1/3}$ (dashed); the
				fitted $L$ exponents are $-0.663\pm0.001$ at every $m_0$.  (b)
				$W_{\rm edge}$ versus $m_0$ at fixed $L=100,200,400,1000$; the law
				agrees to within $1.2$ percent over the entire grid, including
				$m_0=2t$.}
			\label{fig:wedge}
		\end{figure}

		The upturn height obeys
		\begin{equation}
			W_{\rm edge}\simeq 0.74\,\big[t\,(m_0Q)^2\big]^{1/3}
			=0.47\,(\kappa v_x)^{2/3}\,t^{-1/3},
			\label{eq:powerlaw}
		\end{equation}
		with $Q=2\pi/L$ and $\kappa=m_0Q$ the mass slope at the zeros,
		verified against exact
		diagonalization over $L=100\ldots1000$ and $m_0=0.2t\ldots2t$
		(Fig.~\ref{fig:wedge}).  The origin of the exponent becomes explicit in
		the envelope equation obtained by expanding the folded Hamiltonian
		about the saddle point $(\pi,0)$, where both valley band edges touch
		$E=0$ quadratically:
		\begin{equation}
			\big[t\big(-\partial_x^2+k_y^2\big)\sigma_3+\kappa x\,\sigma_2\big]\psi
			=E\psi.
			\label{eq:airy}
		\end{equation}
		This equation holds in the zone-edge boundary layer: the quadratic
		term is the saddle-point curvature at $k_y$ close to $0$ or $\pi$, and
		we use it only to extract the exponent, not as an equation to be solved
		exactly.
		The only dimensionful balance of this problem, $t/x^2\sim\kappa x$,
		selects $x\sim(t/\kappa)^{1/3}$ and $E\sim(t\kappa^2)^{1/3}$, the same
		linear-ramp-versus-quadratic-dispersion balance that underlies Airy
		asymptotics; this is the sense in which the exponent is Airy-type, and
		a simple two-level anticrossing would instead produce integer powers of
		$m_0$.  We stress that Eq.~\eqref{eq:powerlaw} is a scaling law, not an
		exact solution of Eq.~\eqref{eq:airy}: the exponent $2/3$ is fixed by
		this balance, while the prefactor $0.74$ is purely numerical.  The
		fitted $L$ exponents are $-0.663\pm0.001$ at
		every $m_0$ (theory: $-2/3$), and deviations from
		Eq.~\eqref{eq:powerlaw} remain below $1.2$ percent over the entire
		grid, including $m_0=2t$, where the texture is no longer smooth on the
		scale of the wall width.
		The boundary layer also narrows in $k_y$ as $L$ grows.  Consequently,
		in the thermodynamic limit the central band is exactly flat except over
		a boundary region of vanishing height and width; $W_{\rm edge}$ is a
		finite-size boundary effect, not an intrinsic Jackiw--Rebbi scale.  At
		small periods the two zero-bias widths cross: for $m_0=0.2t$ one has
		$W_{\rm JR}>W_{\rm edge}$ below $L^\ast\simeq55$, and $L^\ast$ drops
		below the commensurability threshold as $m_0$ increases, so the
		helical regime of Fig.~\ref{fig:bands}(b) is the small-$L$, small-$m_0$
		corner where Eq.~\eqref{eq:wjr} dominates the observable bandwidth.
		
		\subsection{Bias bump: tunneling through the thinned barrier}
		\label{sec:wbump}
		
\begin{figure}[t]
			\centering
			\includegraphics[width=0.9\columnwidth]{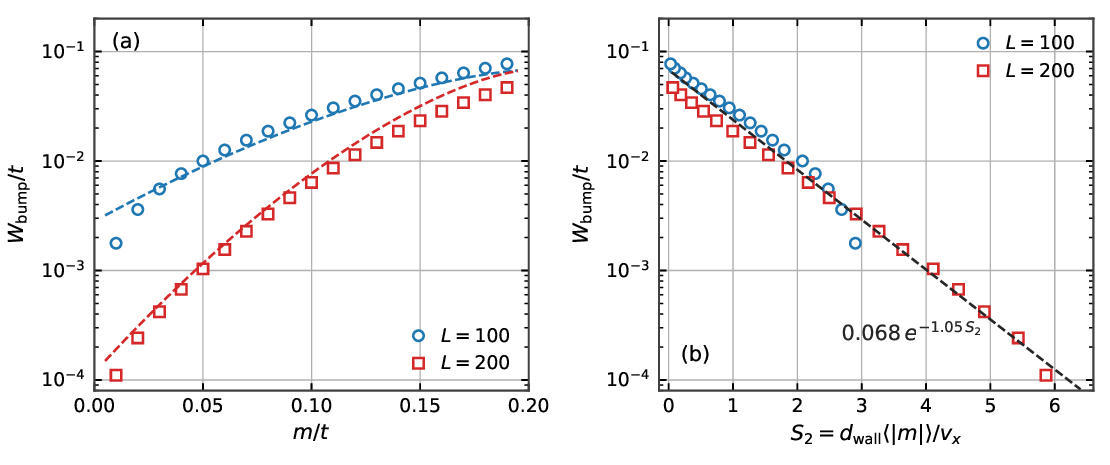}
			\caption{(a) Bias-bump height $W_{\rm bump}$ versus $m$ for
				$m_0=0.2t$ and $L=100,200$ (semilog); dashed lines are the WKB
				estimate $0.068t\,e^{-S_2(m)}$ with $S_2$ from
				Eq.~\eqref{eq:actions}.  (b) Collapse of the same data against $S_2$;
				the dashed fit $W_{\rm bump}=0.068t\,e^{-1.05S_2}$ has an exponent
				within five percent of the WKB value $-1$. }
			\label{fig:wbump}
		\end{figure}

		A uniform bias breaks the half-period antisymmetry $m(x+L/2)=-m(x)$
		that forbids direct hybridization of the two walls at $m=0$, and moves
		the zeros of the period toward each other,
		$d=L\big[1-\theta_0/\pi\big]$ with $\theta_0=\arccos(-m/m_0)$.  The
		region between them has a definite sign of $m(x)$ and acts as a
		tunneling barrier of width $d$ and mean height $\langle|m|\rangle$.
		The resulting mid-zone dispersion of the central band
		[Fig.~\ref{fig:bands}(c)] has height [Fig.~\ref{fig:wbump}(a)]
		\begin{align}
			&W_{\rm bump}\simeq A(m)\,e^{-S_2},
			\nonumber\\
			&S_2=\frac{2\big[m_0\sin\theta_0-m(\pi-\theta_0)\big]}{Qv_x},
			\label{eq:wbump}
		\end{align}
		with $S_2$ the narrow-barrier action of Eq.~\eqref{eq:actions}. The
		measured decay rate at $m_0=0.2t$, $m=0.1t$ is $0.0115$ per site,
		against $S_2/L=0.0109$; the $L=100$ and $L=200$ data collapse onto the
		single curve $W_{\rm bump}\simeq0.068t\,e^{-1.05S_2}$
		[Fig.~\ref{fig:wbump}(b)], whose exponent is within five percent of the
		WKB value $-1$.  At fixed $L$ the bump grows faster than
		quadratically with $m$ (from $2.4\times10^{-4}t$ at $m=0.02t$ to
		$2.3\times10^{-2}t$ at $m=0.15t$ for $L=200$, $m_0=0.2t$), driven by
		the simultaneous thinning and lowering of the barrier.  Two
		consequences follow.  First, the bias response is mesoscopic: at
		$L=10^3$ the same bias leaves a bump below $10^{-6}t$, invisible
		against $W_{\rm edge}\simeq8\times10^{-3}t$, so the thermodynamic-limit
		flat band is robust against bias.  Second, the bias gaps the helical
		nodes by an amount set by the dimerization $||t_1|-|t_2||$, likewise
		exponential in $L$; exact zero-energy crossings survive only at $m=0$,
		where they are protected by the boundary phase of
		Sec.~\ref{sec:helical}.

		Sharp textures behave differently. For a square-wave or short-tanh mass
		profile, the wall slope is held fixed as $L$ grows, the isolated-wall
		approximation becomes asymptotically exact, and the exponential WKB law
		is recovered. On the lattice, sharp walls additionally bind
		intragap lattice states at energies
		\begin{equation}
			E_w=\sqrt{t^2+m_0^2}-t,
			\qquad E_w(E_w+2t)=m_0^2,
			\label{eq:ew}
		\end{equation}
		which are independent of $L$ and of the wall separation
		(Sec.~\ref{sec:lattice}). These states coexist with, and can be mistaken
		for, the Jackiw--Rebbi zero modes; they are absent for smooth textures
		at small $m_0$.

		\section{Valley doubling and the bond-order realization}
		\label{sec:lattice}
		
		The model \eqref{eq:h0} requires no lattice-scale texture of any kind:
		the only ingredients are two opposite-sign lattices and a spatially
		patterned inter-state coupling.  In the cold-atom realization of
		Sec.~\ref{sec:experiment} the two orbitals are simply two hyperfine
		states, which makes Eq.~\eqref{eq:h0} the natural experimental
		platform.
		
		The essential new feature is degeneracy doubling. Because the
		intra-orbital dispersion has two degeneracy points $\pm k_x^0$ for each
		$k_y$, each mass wall binds \emph{two} Jackiw--Rebbi zero modes, one per
		cone. For a single-harmonic texture, which contains only the Fourier
		components $\pm Q$, there is no matrix element connecting the two cones:
		the two modes per wall remain exactly degenerate and exactly pinned at
		zero energy, independently of $L$ and of the bias. Inter-cone
		hybridization requires higher harmonics of the texture (sharp walls) or
		additional scattering (disorder, edges), which converts the exact zeros
		into the narrow dispersive doublets discussed in
		Sec.~\ref{sec:scaling}. This decoupling also explains the lattice result
		that a square-wave texture at $k_y=\pi/2$ supports only the pair
		$\pm E_w$ of Eq.~\eqref{eq:ew} and no zero modes: the sharp walls bind
		lattice states, while the zero-mode channels of the two cones do not
		hybridize and are gapped by the abrupt mass reversal. For a smooth tanh
		wall of width $w=2$ sites, by contrast, exact diagonalization gives
		low-energy states at $\pm5.2\times10^{-4}t$, independent of $L$: these
		are the hybridized Jackiw--Rebbi pair in the regime where the WKB
		exponential applies.
		
	We now make precise
		the relation between the on-site model [Eq. \eqref{eq:h0}] and the
		single-orbital bond-order realization of Sec.~\ref{sec:model}: a
		$(\pi,\pi)$ modulation of the $x$-oriented bonds,
		\begin{equation}
			\delta t_{i,i+\hat x}=\tfrac12\,m(x_i)\,(-1)^{x_i+y_i},
			\label{eq:cbo}
		\end{equation}
		folds into the reduced-zone spinor form of Sec.~\ref{sec:model}, but
		the bond modulation endows the folded mass with a form factor: for
		uniform $m$ the exact folded Hamiltonian is $\epsilon(\mathbf
		k)\sigma_3+m\sin k_x\,\sigma_2$
		[in place of $m\,\sigma_2$ in Eq.~\eqref{eq:hfold}], and for a slow
		texture the form factor is evaluated at the midpoint momentum. The same
		form factor appears in the two-orbital model if the hybridization is
		placed on the bonds in the Qi--Hughes--Zhang form
		$v(x)\,c_n^\dagger(-i\sigma_2)c_{n+\hat x}+\mathrm{H.c.}$
		\cite{qi2008}: the relabeling [Eq. \eqref{eq:map}] maps this term onto
		Eq.~\eqref{eq:cbo} exactly, with $v(x)=m(x)/2$. The bond and on-site
		realizations therefore differ only by the factor $\sin k_x$. On the
		degeneracy line [Eq. \eqref{eq:line}] this factor equals $\sin k_y$ --- the
		same factor carried by the cone velocity $u_x(k_y)=2t\sin k_y$ --- so
		it drops out of the barrier actions [Eq.~\eqref{eq:actions}] and
		reduces to unity at the nodal point, with corrections $O((2\pi n/L)^2)$
		at the $n$-th node. All nodal results of this paper are thus common to
		both realizations; we have verified by exact diagonalization of the two
		lattice models that the helical height $W_{\rm JR}$ and the bias bump
		$W_{\rm bump}$ agree within a few percent over the full range of
		periods. The realizations differ only away from the nodes: the form
		factor vanishes at the van Hove points, so the bond-ordered variant
		leaves them gapless and its zone-edge upturn $W_{\rm edge}$ is
		suppressed by more than an order of magnitude --- the bond-ordered
		system is flatter still than the on-site numbers quoted in
		Sec.~\ref{sec:scaling}. Throughout, $m(x)$ denotes the on-site
		hybridization amplitude, equal to twice the bond-modulation amplitude
		of the equivalent bond texture. The conclusions are unchanged if the
		modulation extends to the $y$-oriented bonds, $\delta t_{i,i+\hat
			y}=\frac12 m(x_i)(-1)^{x_i+y_i}$: the folded mass then acquires on the
		degeneracy line an additional factor $\sin k_y$, again the factor
		carried by $u_x(k_y)$, while the zero modes stay pinned to the mass
		zeros and the mid-zone physics is unchanged.

		\section{Numerical verification}
		\label{sec:numerics}
		
		We diagonalize the model Hamiltonian [Eq.~\eqref{eq:h0}], in the folded
		form Eq.~\eqref{eq:hfold}, on
		supercells of $L$ sites along $x$ with periodic boundary conditions, for
		textures $m(x)=m+m_0\cos(2\pi x/L)$ and sharp-wall variants. Unless
		stated otherwise, $t=1$ sets the energy unit. The figures below combine
		representative spectra at moderate periods, where individual minibands
		are resolvable, with scaling data spanning $L=100\ldots1000$ and
		$m_0=0.2t\ldots2t$, obtained with a dense eigensolver on the
		$2L\times2L$ folded Hamiltonian at fixed $(k_y,k_x^{\rm sc})$.  As a
		reference for the flatness achieved, at $L=1000$, $m_0=0.2t$, $m=0$
		the central band is pinned at zero around $k_y=\pi/2$ to within
		$10^{-15}t$, the first Dirac ladder rung sits at
		$E_1\simeq0.070t$ in agreement with $E_n\simeq\sqrt{8\pi m_0 n/L}\,t$,
		and the full zone-edge upturn is $W_{\rm edge}\simeq8.7\times10^{-3}t$.
		
		For the texture of Fig.~\ref{fig:schematic} the barrier widths are
		known in closed form: at zero bias $d_1=d_2=L/2$ only at $m=0$; for
		the cosine texture $d_{1,2}=(2\pi\mp2\theta_0)/Q$ with
		$\theta_0=\arccos(-m/m_0)$, which at small bias gives
		$d_1\simeq L/2-mL/(\pi m_0)$: the zeros approach each other linearly
		in $m$.  Fig.~\ref{fig:schematic}(b) resolves the zero mode itself: the density from
		exact diagonalization at $k_y=\pi/2$, $k_x^{\rm sc}=0$ (valley
		components $|\psi_a|^2$, $|\psi_b|^2$ and the smoothed total) tracks
		the analytic Jackiw--Rebbi envelope of Eq.~\eqref{eq:jr} with no
		adjustable parameter, confirming that the lattice bound state has the
		continuum profile.  The mode sits at $E=5.2\times10^{-4}t$ rather
		than zero: the residual splitting is the direct hybridization of the
		two walls of the period, the bump scale of Sec.~\ref{sec:wbump}.
		
		Figure~\ref{fig:bands} shows the three regimes of the low-energy
		spectrum as a function of $k_y$.  Panel (a): at $L=1000$, $m_0=0.2t$,
		$m=0$ the central band is flat and pinned at zero to within
		$10^{-15}t$ around $k_y=\pi/2$; the flatness is the chiral pinning of
		Sec.~\ref{sec:jr} combined with the wall-to-wall tunneling $e^{-S_0}$
		[Eq.~\eqref{eq:s0}] suppressed below the solver floor.  The band is
		separated from a ladder of dispersive minibands whose first rung obeys
		$E_n\simeq\sqrt{8\pi m_0 n/L}\,t$, and the only visible dispersion is
		the zone-edge upturn analyzed in Sec.~\ref{sec:wedge}.  Panel (b): at
		$L=100$ the same texture produces the helical miniband around zero
		energy, oscillating through nodes evenly spaced by $2\pi/L$; the
		oscillation height is $W_{\rm JR}$ of Eq.~\eqref{eq:defjr}, now that
		the tunneling is appreciable at this period, and the node positions
		are the boundary-phase condition of Sec.~\ref{sec:helical}, with the
		node at $k_y=\pi/2$ exact for this $L\equiv0\ (\mathrm{mod}\ 4)$
		period.  Panel (c): a bias $m=0.1t$ at $L=200$ lifts the mid-zone
		into the bump analyzed in Sec.~\ref{sec:wbump}, the dispersion of the
		dimerized SSH ring of Sec.~\ref{sec:ssh}, while the zone-edge upturn
		is essentially unchanged.  All three panels mark the zone-edge height
		$W_{\rm edge}$, measured from $E=0$, to emphasize that it is present
		in every regime.
		
		Figure~\ref{fig:wjr} tests the intrinsic hybridization scale: the
		helical oscillation height $W_{\rm JR}$ [Eq.~\eqref{eq:defjr}], sampled
		at the antinode positions $k_y=\pi/2+n\pi/L$ so that both node
		parities are covered, follows the WKB exponential [Eq. \eqref{eq:wjr}] over
		five to ten decades, with decay rates within four percent of the
		barrier action $m_0L/\pi v_x$.  Data points terminate where $W_{\rm JR}$ sinks
		below the $\sim10^{-13}t$ noise floor of the eigensolver; the
		measurable window shrinks with $m_0$ as $L\lesssim28\pi v_x/m_0$. Figure~\ref{fig:wjr}(b) confirms the $m_0$ dependence at fixed $L$: the data follow
		the fixed theoretical slope $-L/\pi v_x$, leaving the prefactor
		$C(m_0)$ as the only free parameter.
		
		Figure~\ref{fig:wedge} tests the zone-edge width.  On log-log scales
		$W_{\rm edge}$ follows the power law [Eq. \eqref{eq:powerlaw}] in both $L$
		(slope $-2/3$) and $m_0$ (slope $2/3$) over the full grid, with
		deviations below $1.2$ percent everywhere, including $m_0=2t$ where
		the texture is no longer smooth on the scale of the wall width.  This
		is the width that any max-min
		bandwidth measurement returns at large $L$.

		Figure~\ref{fig:wbump} quantifies the bias response.
		Figure~\ref{fig:wbump}(a) shows the bump height
		[Eq.~\eqref{eq:defbump}] versus $m$ for $L=100$ and $200$: faster than
		quadratic growth at fixed $L$, and exponential suppression as $L$
		grows.  Figure~\ref{fig:wbump}(b) collapses the data onto the
		narrow-barrier action $S_2$ of Eq.~\eqref{eq:actions}, confirming that
		the bump is the tunneling transparency of the thinned barrier,
		$W_{\rm bump}\propto e^{-S_2}$; at $L=10^3$ the bump is below
		$10^{-6}t$ for any $m<m_0$.

		\section{Experimental realizations}
		\label{sec:experiment}
		
		The ingredients required by Eqs.~\eqref{eq:h0}--\eqref{eq:mx} --- two
		opposite-sign square lattices and a spatially patterned on-site
		inter-orbital coupling with a tunable offset --- are available in
		several platforms.
		
		\emph{Cold atoms.} The model
		Eq.~\eqref{eq:h0} maps directly onto standard optical-lattice
		ingredients. Two hyperfine states $|a\rangle,|b\rangle$ of a fermionic
		species encode the two orbitals; each is loaded into its own square
		lattice, and the orbital-polarized condition $\epsilon_b=-\epsilon_a$
		is the opposite sign of the two tunnelings, set by the lattice phases
		or by Floquet shaking. A two-photon Raman coupling between the
		hyperfine states supplies the inter-orbital hopping, and patterning its
		intensity with a digital micromirror device or a spatial light
		modulator imprints $m(x)=m+m_0\cos(Qx)$ directly, with no lattice-scale
		texture required. The uniform bias $m$ is the unmodulated part of
		the Raman coupling, controlled by the two-photon detuning. The
		parameter demands are modest: for a typical lattice constant
		$a\simeq0.5\,\mu$m the modulation period $La$ is $5$ to $50\,\mu$m for
		$L=10$ to $100$, well above the few-micron resolution of current
		digital-micromirror or spatial-light-modulator optics; the Raman Rabi
		frequency sets $m_0$ against the tunneling, and the temperature must
		resolve the width of interest, $k_BT<W_{\rm edge}$ in the flat-band
		regime versus $k_BT<W_{\rm JR}$ for the helical oscillations, the
		latter tunable through $L$ and $m_0$ [Eq.~\eqref{eq:wjr}]. The required
		capabilities have been demonstrated piecemeal: fully tunable hopping
		\cite{tarruell2012}, topological pumping of Rice--Mele domain-wall
		states \cite{lohse2018}, edge-state imaging in SSH superlattices
		\cite{atala2013}, interacting SSH chains of Rydberg atoms
		\cite{deleseleuc2019}, and moir\'e optical lattices \cite{meng2023}.
		The single-orbital bond-order variant of
		Sec.~\ref{sec:lattice} is also feasible, at the price of an
		additional short-period superlattice generating the $(\pi,\pi)$ bond
		texture; its nodal physics is identical, with an even flatter central
		band. Detection is the strongest asset of this platform:
		quantum-gas microscopy resolves the zero-mode wavefunctions
		[Eq. \eqref{eq:jr}] in real space, doubly degenerate at each wall in the
		two-orbital model; band-mapping or momentum-resolved spectroscopy
		resolves the flat-to-helical crossover versus $L$; and the
		dimerization $t_1/t_2=e^{-mL/u_x(k_y)}$ can be read off from the miniband
		splitting, providing a direct test of Eq.~\eqref{eq:dimer}.
		
		\emph{Kekul\'e-ordered graphene.} A Kekul\'e bond texture folds the
		graphene Dirac cones and opens a gap whose sign is a synthetic mass
		\cite{hou2007};
		ordered molecular monolayers on Cu(111) have realized this order and its
		topology directly \cite{bao2021}. A spatially patterned Kekul\'e
		amplitude, imposed by a moir\'e or by patterned adsorbates, realizes
		$m(x)$ with Jackiw--Rebbi modes at its zeros; scanning tunneling
		spectroscopy maps both the zero-mode localization and the bias response.
		
		\emph{STM designer lattices.} Atomic assembly on Cu(111) has delivered
		designer electronic lattices with programmed hopping patterns
		\cite{drost2017,slot2017}. A $(\pi,\pi)$ bond texture with a slow envelope --- the bond-order
		realization of Sec.~\ref{sec:lattice} --- can be
		patterned directly as a sequence of molecule positions, and local
		spectroscopy gives energy- and position-resolved access to the minibands
		of Fig.~\ref{fig:bands}.
		
		\emph{Bilayer-graphene domain-wall networks.} Stacking domain walls in
		bilayer graphene bind topologically protected one-dimensional modes
		\cite{martin2008,ju2015}. A periodic array of such walls with controlled spacing is
		a close analog of our mass-wall lattice, and the wall separation,
		analogous to $d_1,d_2$, is tunable by strain or gating.
		
		\emph{Microwave and photonic arrays.} Coupled dielectric resonators have
		realized SSH physics and its edge states with site-resolved access
		\cite{bellec2013,ozawa2019}. The mass texture is encoded in the resonator
		couplings, and the node ladder [Eq. \eqref{eq:nodes}] appears as a sequence of
		transmission zeros as the synthetic $k_y$ is swept.
		
		Across these platforms the sharpest signatures are (i) the flat,
		chirally pinned band whose residual width is the algebraic zone-edge
		upturn [Eq. \eqref{eq:powerlaw}], (ii) the node ladder $L\cos k_y=2\pi n$ in
		the helical regime, with its oscillation height governed by the WKB
		exponential Eq. \eqref{eq:wjr}, and (iii) the exponential bias dimerization
		$t_1/t_2=e^{-mL/u_x(k_y)}$, observable as the mesoscopic mid-zone bump
		[Eq. \eqref{eq:wbump}] before it switches off exponentially in $L$.
		
		\section{Discussion}
		\label{sec:discussion}
		
		\emph{Robustness of the pinning.} The exact zero-energy pinning of the
		central miniband relies on the lattice chiral symmetry $\Gamma$
		(Sec.~\ref{sec:topology}).  A further-neighbor intra-orbital hopping
		$t'$ (of the same sign for the two orbitals) or a staggered on-site
		potential breaks $\Gamma$ and shifts the zero modes by an
		energy of order the perturbation strength, converting the exactly flat
		band into a dispersive one of width set by $t'$ rather than by
		$W_{\rm JR}$.  The consequences of breaking $\Gamma$ are themselves
		spectroscopically structured: a spatially uniform symmetry-breaking
		term shifts the central miniband rigidly, moving the helical crossings
		off zero energy without gapping them, whereas a term modulated on the
		scale of the texture gaps the nodes selectively; a $\Gamma$-preserving
		perturbation, such as a distortion of the envelope or a longer-range
		hopping along the wall, leaves the pinning intact and only renormalizes
		$t_1/t_2$ and the node positions.  The loss of pinning is therefore not
		a single failure mode, and the three cases are distinguishable in a
		band map.  Since $W_{\rm JR}$ is exponentially small, even weak
		symmetry breaking dominates the intrinsic width.  The experimentally
		relevant statement is therefore the reverse one: as long as the
		symmetry-breaking scale stays below the zone-edge width $W_{\rm edge}$
		[Eq.~\eqref{eq:powerlaw}], the central band looks flat everywhere
		except in the narrowing zone-edge boundary layer, and the
		flat-to-helical crossover remains observable.  The smooth cosine shape
		of the texture is likewise inessential: any single-harmonic envelope
		with simple, well-separated zeros gives the same fixed-$k_y$ Dirac
		structure, while sharp walls add the lattice bound states of
		Eq.~\eqref{eq:ew} as a spectroscopic side band.
		
		\emph{Interaction scales.} The width hierarchy provides the energy
		scales at which interactions become relevant.  In the flat-band regime
		the residual bandwidth is the algebraic $W_{\rm
			edge}\sim[t(m_0Q)^2]^{1/3}$ [Eq.~\eqref{eq:powerlaw}]; a Hubbard-scale interaction exceeding it places
		the half-filled central band in the strongly correlated regime
		\cite{cao2018,andrei2020}, with
		the symmetry-pinned zero modes as the active degrees of freedom.  In
		the helical regime the relevant scale is instead $W_{\rm JR}$, tunable
		over decades through $L$ and $m_0$ [Eq.~\eqref{eq:wjr}], while the bias
		provides an independent knob through $W_{\rm bump}$
		[Eq.~\eqref{eq:wbump}] and through the $\mathbb{Z}$ dimerization
		[Eq. \eqref{eq:dimer}] of the SSH ring.
		
		\emph{Uses of the construction.} Beyond the specific model, the
		fixed-$k_y$ reduction is a design tool.  Because a uniform bias enters
		the SSH ring only through the tunneling ratio [Eq. \eqref{eq:dimer}],
		exponentially amplified by the period, a weak modulation switches the
		central band between the pinned flat, helical, and dimerized regimes
		and flips the $\mathbb{Z}$ invariant: a narrow band with a built-in,
		exponentially sensitive topological switch.  The three-width
		decomposition of Sec.~\ref{sec:scaling} in turn makes spectroscopy
		diagnostic: the node ladder, the exponential rates of $W_{\rm JR}$ and
		$W_{\rm bump}$, and the power law of $W_{\rm edge}$ are independent
		fingerprints that identify which mechanism a measured bandwidth
		actually reflects, in any of the platforms of Sec.~\ref{sec:experiment}.
		Finally, the reduction itself transfers: any exactly nested
		two-dimensional dispersion with a slowly modulated valley-mixing mass
		admits the same Dirac-ring description, and the bond-order
		realization of Sec.~\ref{sec:lattice} shows that the construction
		transfers to single-orbital systems.

		\emph{Open directions.} We have taken the mass texture to be
		independent of $y$ throughout [Eq.~\eqref{eq:mx}], so that $k_y$
		remains a good quantum number and the fixed-$k_y$ reduction applies;
		the Jackiw--Rebbi states are then one-dimensional, bound to the lines
		$m(x)=0$. Modulating the envelope
		along $y$ as well, $m(x)\to m(x,y)$, is more radical: $k_y$ is no
		longer conserved, the mass zeros form a two-dimensional network rather
		than parallel lines, and the states bound to it define a real-space
		Jackiw--Rebbi superlattice of discrete, coupled zero modes. Each link
		of the network is still governed by the wall-and-barrier mechanism
		analyzed here, and we leave this extension to future work. Within the
		one-dimensional setting itself, several further extensions follow
		naturally from the fixed-$k_y$ reduction. Treating the texture
		self-consistently would let the flat band feed back on the texture
		itself, a question raised already for field-stabilized bond and
		excitonic textures.
		Envelopes with several harmonics reconnect the two valleys and
		interpolate toward the sharp-wall limit, where the intragap states
		[Eq. \eqref{eq:ew}] appear. Disordered or quasiperiodic envelopes randomize
		the wall positions and hence the effective SSH ring, a controlled
		route to random-dimer physics. Finally, the
		node ladder [Eq. \eqref{eq:nodes}] and the dimerization law
		[Eq. \eqref{eq:dimer}] are sharp, quantitative benchmarks against which any
		experimental realization of Sec.~\ref{sec:experiment} can be tested.
		
		\section{Summary}
		\label{sec:summary}
		
		We have shown that a two-orbital square-lattice model with
		opposite-sign intra-orbital hopping and a slowly modulated on-site
		inter-orbital hybridization realizes, independently at
		each transverse momentum $k_y$, a Dirac Hamiltonian with a periodic
		mass, unifying three regimes: chirally pinned Jackiw--Rebbi bands that become flat in the large-period limit,
		helical minibands whose nodes are enforced by the boundary phase
		$\Phi=[k_x^{\rm sc}+\cos k_y]L$, and a bias-dimerized SSH ring with
		tunneling ratio $e^{-mL/u_x(k_y)}$ and a winding number set by the sign of
		the bias. These results rest on the fixed-$k_y$ Dirac reduction; its
		lattice corrections, the van Hove cutoff and the parity refinement of
		the node positions, are quantified in Secs.~\ref{sec:model} and
		\ref{sec:helical} and tested against exact diagonalization in
		Sec.~\ref{sec:numerics}. The central miniband carries three decoupled
		widths: the helical oscillation height $W_{\rm JR}\simeq
		C(m_0)e^{-m_0L/\pi v_x}$, which follows the WKB exponential with
		measured rates within four percent of the barrier action; the zone-edge
		upturn $W_{\rm edge}\simeq0.74[t(m_0Q)^2]^{1/3}$, which sets the
		max-min bandwidth at large $L$; and the bias-induced mid-zone bump
		$W_{\rm bump}\propto e^{-S_2}$, a mesoscopic effect that switches off
		exponentially in $L$. Sharp textures instead bind lattice states at
		$E_w=\sqrt{t^2+m_0^2}-t$; the second valley cone doubles the zero modes
		to two per wall, and the bond-order realization of
		Sec.~\ref{sec:lattice} shares the nodal physics while suppressing the
		zone-edge upturn.
		
		The most distinctive testable predictions are the boundary-phase node
		ladder and the bias-controlled dimerization law
		$t_1/t_2=e^{-mL/u_x(k_y)}$: both are sharp spectral criteria, readable from
		a band map or from transmission zeros, and both are within reach of the
		experimental platforms detailed in Sec.~\ref{sec:experiment}. More
		broadly, the construction provides a minimal, analytically tractable
		reference model in which flat-band formation, domain-wall
		hybridization, and topological switching can be computed quantitatively
		and benchmarked experimentally; the extensions mapped out in
		Sec.~\ref{sec:discussion} make it the starting point for a family of
		problems rather than a single answer. The natural next question is the
		fate of the pinned band at half filling
		under interactions, where the exponentially tunable $W_{\rm JR}$ sets
		the scale at which a Hubbard-scale repulsion enters the strongly
		correlated regime.
		
		\begin{acknowledgments}
			This work was supported by the 
			Guangdong Provincial Quantum Science Strategic Initiative (GDZX2601002).
		\end{acknowledgments}

	\bibliography{ref}
\end{document}